\documentclass[aps,prl,reprint,floatfix,superscriptaddress]{revtex4-1}
\usepackage{amsmath}
\usepackage{amssymb}
\usepackage{bm}
\usepackage{epsfig}
\usepackage{graphicx}
\usepackage{color}
\usepackage[colorlinks=true]{hyperref} 
\usepackage[utf8]{inputenc}
\usepackage[english]{babel}

\begin{document}
\title{Nanosecond-scale magneto-exciton energy oscillations in quantum wells}
\author{A.~V.~Trifonov}
\email[correspondence address: ]{a.trifonov@spbu.com}
\affiliation{Spin Optics Laboratory, St. Petersburg State University, St. Petersburg, 198504, Russia
}


\author{E. S. Khramtsov}
\affiliation{Spin Optics Laboratory, St. Petersburg State University, St. Petersburg, 198504, Russia
}
\author{K. V. Kavokin}
\affiliation{Spin Optics Laboratory, St. Petersburg State University, St. Petersburg, 198504, Russia
}
\affiliation{A. F. Ioffe Physical-Technical Institute, Russian Academy of Sciences, St Petersburg 194021, Russia
}
\author{I. V. Ignatiev}
\affiliation{Spin Optics Laboratory, St. Petersburg State University, St. Petersburg, 198504, Russia
}
\author{ A. V. Kavokin}
\affiliation{Spin Optics Laboratory, St. Petersburg State University, St. Petersburg, 198504, Russia
}
\affiliation{Institute of Natural Sciences, Westlake University, No.18, Shilongshan Road, Cloud Town, Xihu District, Hangzhou, China
}
\affiliation{CNR-SPIN, 1, Viale del Politecnico, I-00133, Rome, Italy
}
\author{Y. P. Efimov}
\affiliation{Resource center "Nanophotonics", St. Petersburg State University, St. Petersburg, 198504, Russia
}
\author{ S. A. Eliseev}
\affiliation{Resource center "Nanophotonics", St. Petersburg State University, St. Petersburg, 198504, Russia
}
\author{P. Yu. Shapochkin}
\affiliation{Resource center "Nanophotonics", St. Petersburg State University, St. Petersburg, 198504, Russia
}
\author{M. Bayer}
\affiliation{Experimentelle Physik 2, Technische Universittat Dortmund, D-44221 Dortmund, Germany
}
\affiliation{A. F. Ioffe Physical-Technical Institute, Russian Academy of Sciences, St Petersburg 194021, Russia
}

\date{\today}

\begin{abstract}
We report on the experimental evidence for a nanosecond time-scale spin memory based on nonradiative excitons. The effect manifests itself in magnetic-field-induced oscillations of the energy of the optically active (radiative) excitons. The oscillations detected by a spectrally-resolved pump-probe technique applied to a GaAs/AlGaAs quantum  well structure in a transverse magnetic field persist over a time scale, which is orders of magnitude longer than the characteristic decoherence time in the system. The effect is attributed to the spin-dependent electron-electron exchange interaction of the optically active and inactive excitons. The spin relaxation time of the electrons belonging to nonradiative excitons appears to be much longer than the hole spin relaxation time.
\end{abstract}


\maketitle

Excitons are crystal quasiparticles that can be generated by light and that may eventually recombine emitting light~\cite{Gross1952, Excitons}. As such, they are promising for storing the optically encoded information and keeping memory of the intensity, phase, and polarization of light. Applications of excitons for optical storage are limited by their short radiative lifetime (typically, on the order of 10 -- 100 ps) and even shorter coherence time (on the order of a few picoseconds). Nonradiative, also referred to as dark or optically inactive, excitons that are decoupled from light due to the specific selection rules for optical transitions are widely discussed as the most promising exciton memory agents~\cite{Combescot-PRL2007, Combescot-PRL2011, Gantz-PRB2016, Rapaport,  JBloch}. They possess lifetimes on a nanosecond or longer scale and affect many processes in  optically excited quantum wells (QWs) ~\cite{Feldmann-PRL1987, Honold-PRB1989, Damen-PRB1990, Deveaud-PRL1991, Szczytko-PRL2004, Trifonov-PRB2015}, quantum dots~\cite{Gershoni-NatPhys2010, Gershoni-PRX2015}, microcavities~\cite{Glazov-PRB2009, Wouters-PRB2013, Giorgi-PRL2014, Kavokin-book2017}, and 2D-materials~\cite{Berg-PRB2018, Malic-PRMat2018}. On the other hand, a rapid thermalization of the reservoir of nonradiative excitons usually leads to the loss of coherence on a few-picosecond scale. 

\begin{figure}[htp]
\centering
\includegraphics[width=1\linewidth]{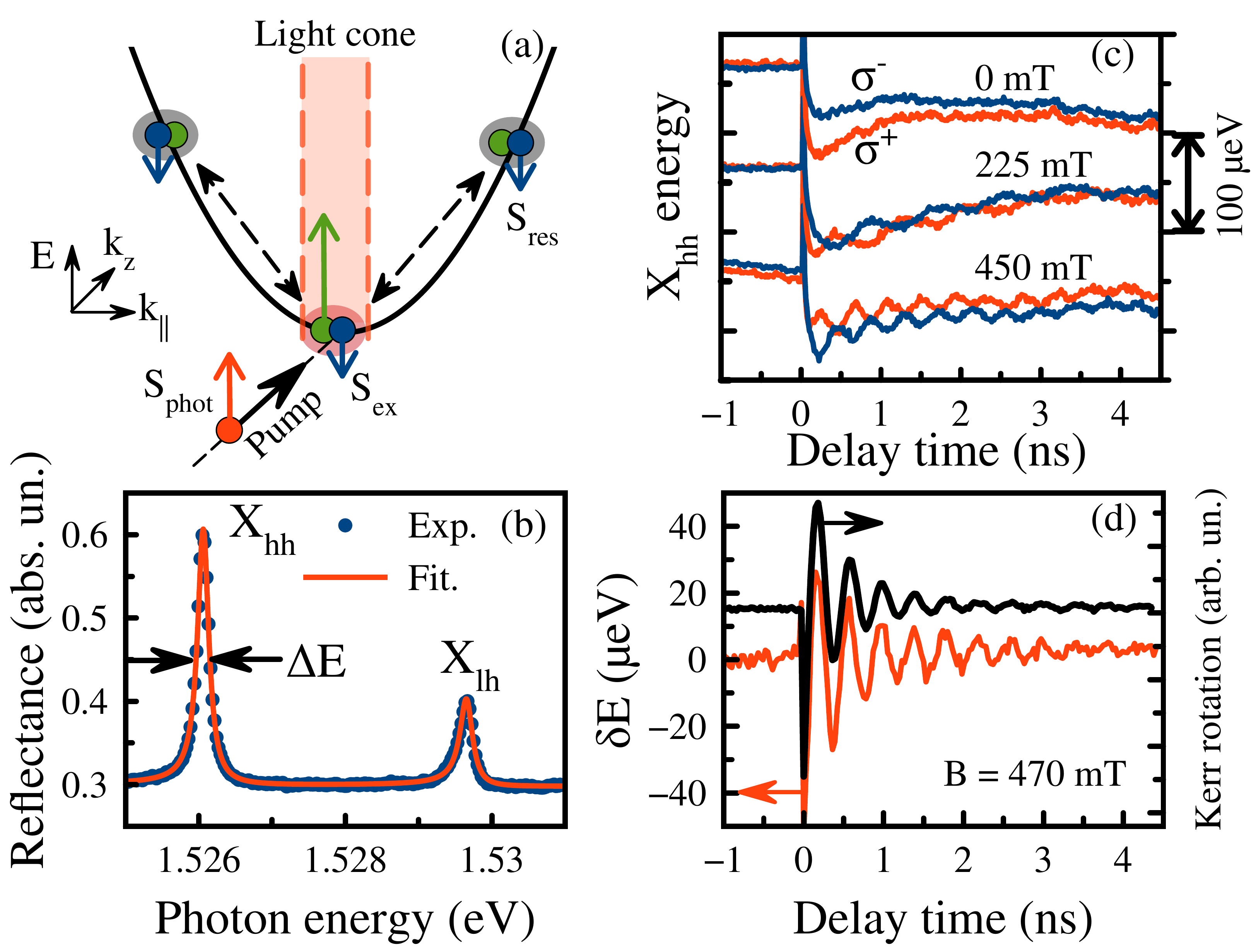}
\caption{(a) A scheme illustrating optically active and inactive excitons and the exchange interaction of spins of the radiative excitons within the light cone with the spin-polarized excitons in the nonradiative reservoir. (b) Reflectivity spectrum of the 14-nm GaAs/AlGaAs QW (blue dots). The red dashed line shows the fit of the exciton resonances by Eqs.~(\ref{eqn1}, \ref{eqn2}). $\Delta E \approx 180$~$\mu$eV. (c) Time evolution of the Xhh exciton energy in the transverse magnetic field in the $\sigma^+$ (red curves) and $\sigma^-$ (blue curves) polarizations. The magnetic field magnitudes are indicated near each pair of curves. The curves are shifted for clarity. (d) Comparison of the dynamics of the Kerr rotation signal (black curve) and of the exciton energy splitting (red curve). The sample temperature is $T = 6$~K.}
\label{Fig1}
\end{figure}

The capacity of a reservoir of nonradiative excitons to serve as an optical polarization or spin storage is yet to be fully revealed. Here we study the spin memory effects in the excitonic system by means of time-resolved magneto-optical spectroscopy. In our experiment the reservoir consists of excitons with large in-plane wave vectors strongly exceeding the wave vector of light [see Fig.~\ref{Fig1}(a)] so that the $k$-vector selection rules do not allow these excitons to absorb or to emit light~\cite{comment2}. Nevertheless dark excitons can be optically addressed via their interaction with the optically active (bright) excitons~\cite{Trifonov-PRB2015}. 

We have developed an experimental approach allowing for a direct access to the spin polarization of reservoir excitons. We observe a robust exciton spin polarization lasting several nanoseconds. It manifests itself in magnetic-field-induced oscillations of the optically active exciton energy due to their exchange interaction with the reservoir of spin-polarized nonradiative excitons.  

A high-quality heterostructure with a 14-nm GaAs/Al$_{0.03}$Ga$_{0.97}$As quantum well (QW)  was experimentally studied. The structure was grown by molecular beam epitaxy at the n-doped GaAs substrate. Due to the small content of Al in the barrier layers, their height is relatively small, about 25~meV for electrons and 12~meV for holes. Fig.~\ref{Fig1}(b) shows a reflectance spectrum of the sample in the spectral vicinity of the exciton resonances. The main features observed in the spectrum can be ascribed to optical transitions to the quantum-confined heavy-hole (Xhh) and light-hole (Xlh) exciton states in the QW. The very small spectral widths of the exciton resonances confirm the ultra-high quality of the structure. These resonances can be precisely modeled by a phenomenological theory described in Refs.~\cite{Ivchenko-book, Trifonov-PRB2015}.

Within this model the amplitude reflection coefficient of light from a QW can be written in the form:
\begin{equation}
\label{eqn1}
r_{X}=\frac{i\Gamma_{0}}{\tilde{\omega}_{0}-\omega - i(\Gamma_{NR}+\Gamma_{0})}.
\end{equation}
Here the parameter $\Gamma_{0}$ describes the radiative decay rate of the exciton state, $\Gamma_{NR}$ is the rate of nonradiative relaxation from this state,  $\tilde{\omega}_{0}$ is the frequency of the exciton transition. These three quantities are considered to be fitting parameters of the model. The reflectivity spectrum of the structure is then given by:
\begin{equation}
\label{eqn2}
R=\left|\frac{r_s+r_{X}e^{i\phi}}{1+r_s r_{X}e^{i\phi}}\right|^2,
\end{equation}
where $r_s$ is the amplitude reflection coefficient of the sample surface and $\phi$ is the phase shift of the light travelling from the sample surface to the QW and back.

The good agreement of the experimental and modelled spectra shown in Fig.~\ref{Fig1}(b) indicates that no significant inhomogeneous (Gaussian-like) broadening is present in this structure. This allows us to obtain reliable values of all the fitting parameters. For the Xhh resonance shown in Fig.~\ref{Fig1}(b) the fitting parameters are: $\hbar \Gamma_{0} = 30\pm1$~$\mu$eV, $\hbar \Gamma_{NR} =  45\pm 2 $~$\mu$eV, $E_{Xhh} = \hbar \tilde{\omega}_{0} = 1526.061\pm0.002$~meV. One can see that the energy of the exciton states can be obtained with a high accuracy of about 2~$\mu$eV. This opens the way to highly sensitive experiments for the study of interaction of  photocreated excitons with other quasiparticles in the structure.

We have developed a spectrally-resolved pump-probe experimental technique with the circularly polarised 2-picosecond pump pulse exciting the structure at some spectral point while the spectrally-broad 100-femtosecond probe pulse is used to detect the reflection spectrum at each delay between the pump and probe pulses. The linearly polarized probe beam reflected from the sample is split into two circularly polarized components. Spectra of  both components are simultaneously measured by an imaging spectrometer equipped by a CCD detector. In this way, two reflectance spectra in both circular polarizations are detected. The analysis of the spectra measured at different delays using Eqs.~(\ref{eqn1}, \ref{eqn2}) allows us to obtain the dynamics of the essential excitonic parameters. 

Fig.~\ref{Fig1}(c) shows the dependence of the energy  of the Xhh exciton resonance on the delay between pump and probe pulses. A small magnetic field is applied to the structure perpendicular to the growth axis (Voigt geometry). We see that the exciton energy undergoes an instantaneous jump and rapid decay at small delays followed by a smooth change when no magnetic field is applied. The tail of the exciton energy dynamics, however, becomes oscillating in the presence of the magnetic field. The oscillations are opposite in sign for $\sigma^+$ and $\sigma^-$ polarizations detection. The frequency of the oscillations increases with the magnetic field increase. 

To clarify the origin of these oscillations, we have compared the oscillations in energy splitting, $\delta E= E^{\sigma^{+}}_{{hh}}-E^{\sigma^{-}}_{{hh}}$, with the oscillating Kerr rotation signal measured in the same experimental conditions, see Fig.~\ref{Fig1}(d). One can see that both  signals look very similar. However, the oscillation parameters are different in these two measurements: the Kerr angle and the exciton energy respectively. It is well known~\cite{Crooker-PRB1997, GlazovPSS2012, Dyakonov-book} that the oscillating Kerr signal is determined by the spin polarization that precesses about the magnetic field. Therefore, we may conclude that the energy splitting $\delta E_{hh}$ may also be a result of the exchange interaction of the radiative excitons with some reservoir of polarised spins precessing in the external magnetic field.

\begin{figure}[b]
\centering
\includegraphics[width = 0.95\linewidth]{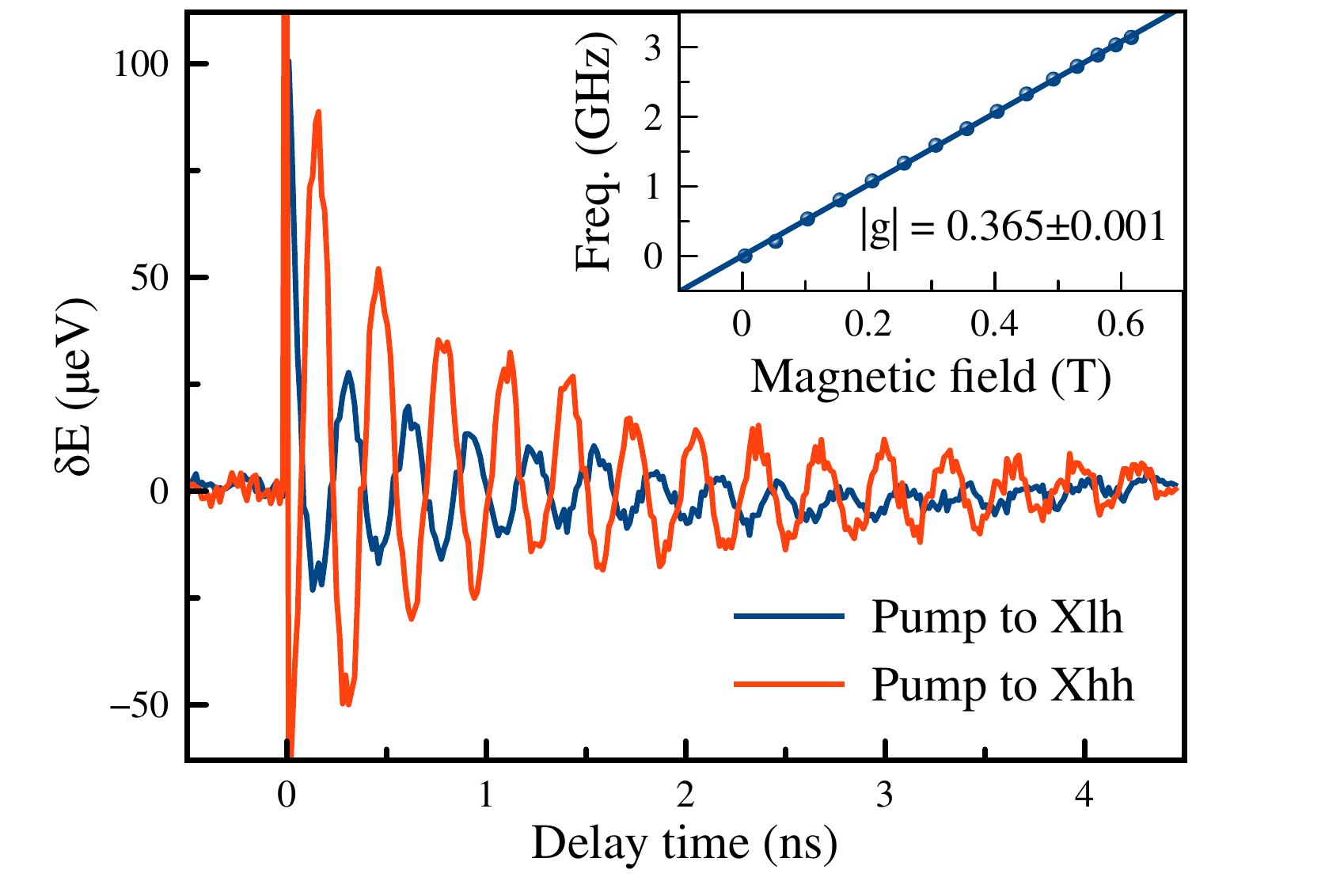}
\caption{The time delay dependence of energy difference  $\delta E= E^{\sigma^{+}}_{{hh}}-E^{\sigma^{-}}_{{hh}}$ of the Xhh exciton energies measured at the $\sigma^+$ and $\sigma^-$ circular polarizations under the  $\sigma^+$ excitation into the Xhh (red curve) and Xlh (blue curve) exciton resonances. Excitation density $P = 50$~W/cm$^2$; magnetic field strength $B = 0.6$~T; sample temperature $T = 5.6$~K. Inset shows the oscillation frequency vs magnetic field strength extracted from the experiment (points) and the linear fit by function $\Omega = (|g| \mu_B B)/\hbar$.
}
\label{Fig2}
\end{figure}

To reveal the physical nature of the spin reservoir that is responsible for the oscillations, we have measured the oscillating energy splitting at different magnitudes of the magnetic field exciting at the heavy-hole and light-hole resonances as shown in Fig.~\ref{Fig2}. One can see that the  phase of the $ \delta E$ oscillations is opposite in the two experiments. This is a signature of the difference in selection rules for the optical excitation of Xhh and Xlh excitons~\cite{Ivchenko-book}. Clearly, the optically active heavy-hole and light-hole excitons created by light with the same helicity of polarization involve electrons with opposite spins. 

The dependence of the oscillation frequency on the applied external magnetic field is shown in the inset of figure~\ref{Fig2}. It is clearly seen that the frequency dependence on the magnetic field is linear. From the slope of this line we can determine the g-factor $ | g | = 0.365 \pm 0.001 $, which nearly coincides with the known value of the electron g-factor in QWs~\cite{Yugova-PRB2007}.

The magnitude of the g-factor and the inversion in the phase of the oscillations upon excitation of the Xlh and Xhh exciton resonances are two key experimental  findings that point to the mechanism of the oscillations. We conclude that the oscillating energy splitting of the exciton levels is due to the exchange interaction of excitons with the long-lived electrons whose spins precess about the applied external magnetic field. 
These can be free resident electrons, photocreated free electrons, or electrons in the excitons of a long-lived nonradiative reservoir. 

To identify the origin of these electrons, we have performed a theoretical estimate of the exchange interaction between the bright excitons and the long-lived electrons as well as of the electron density $n_e$ required to obtain the observed energy shifts. 
The spin Hamiltonian of the exchange interaction reads~\cite{Ciuti-PRB1998,LandauT3}:
\begin{equation}
\hat{H}_S = \Delta_0 \hat{i}_z \hat{s}_z + J_{ee} n_e \left ( \hat{\vec{s}} \cdot \langle \vec{S} \rangle\right),
\label{Ham1}
\end{equation}
Here $\hat{s}_z$ and $\hat{i}_z$ are the projections of the electron and hole spins belonging to the bright exciton on the growth axis $z$, $\Delta_0$ is electron-hole exchange interaction energy, $\langle \vec{S} \rangle$ is the average spin in the reservoir, and $J_{ee}$ is the exchange interaction constant. The explicit expression for $J_{ee}$ and its numerical calculation are given in the Supplementary material~\cite{Suppl}. We have also estimated the constant $\Delta_0$ for the structure under study. The obtained value~\cite{Suppl}, $\Delta_0 < 20$~$\mu$eV, is small compared to the observed exciton energy splitting caused by the interaction of the exciton spin with the reservoir of electron spins. 
 We should also note that the exchange interaction of a hole in the bright exciton with the reservoir electrons  is much weaker and can be neglected~\cite{Ciuti-PRB1998}. 

The diagonalization of a Hamiltonian~(\ref{Ham1}) gives rise to four eigenstates. When the average spin $\langle S \rangle$ is directed along the growth axis $z$, the bright and dark exciton states are not mixed and optical transitions are allowed only to the bright exciton states.
When the reservoir spin is rotated perpendicular to the $z$ axis by the applied magnetic field, the bright and dark exciton states are mixed and all four exciton transition are allowed. All the splittings, however, are much smaller then the exciton line broadening $\hbar(\Gamma_0+\Gamma_{NR})$. Therefore the effect of exchange interaction is observed as a shift of a single exciton resonance when the reservoir spin is rotated. The difference in the energy positions of the single resonance seen in the $\sigma^+$ and $\sigma^-$ polarisations is described by (see eq. (8) in Suppl. Mat.~\cite{Suppl})
\begin{equation}
\delta E = J_{ee} n_e  \langle S_z\rangle,
\label{Eqn4}
\end{equation}
where $\langle S_z\rangle$ is the $z$-projection of the reservoir spin.

Let us first consider the exchange interaction of bright excitons with a  reservoir of free electrons. The correspondence interaction constant is~\cite{Suppl}: $J_{ee}^{xe} = 18 \pm 2$~$\mu$eV$\times \mu^2$. If the reservoir electrons are totally polarized, that is, $\langle S_z\rangle = 1/2$, we obtain from Eq.~(\ref{Eqn4}) the minimum areal electron density, $n_{e} = 1.1 \times 10^9$~cm$^{-2}$ (for $\delta E =100  \mu$eV, see Fig. 2). If the reservoir is composed by resident electrons, their average polarization $\langle S_z\rangle\ll 1/2$ and the required areal density $n_{e} \gg 10^9$~cm$^{-2}$. The electrons with such areal density should give rise to the trion (negatively charged exciton) peaks in the optical spectra~\cite{Shields-PRB1995, Astakhov-PRB2002}. We, however, did not observed such features in both the photoluminescence and reflectance spectra. We should note that the trion peaks are observed in the intentionally n-doped structures~\cite{Astakhov-PRB2002}.
But our structure is undoped. Therefore we may assume that the resident electrons cannot be responsible for the observed effect.

The free electrons in the reservoir could be, in principle, created by optical pumping. However, we have used the resonant pumping into the lowest exciton state which makes this scenario unlikely. Indeed, the electrons are coupled with holes in the photocreated excitons. The Coulomb energy of the coupling, $R_X \approx 7$~meV in the QW under study~\cite{Khramtsov-JAP2016}, which is one order of magnitude larger than the thermal energy, $kT \approx 0.5$~meV. Therefore the photocreation of free electrons is supposed to be a highly inefficient process in our case. 

Eventually we come to the conclusion that the electrons that belong to the reservoir excitons are responsible for the observed energy shifts. The interaction of the bright excitons with those in the nonradiative reservoir via the electron-electron exchange~\cite{Suppl}, $J_{ee}^{xx} = 11.4 \pm 0.8$~$\mu$eV$\times \mu^2$. Therefore the required areal density of the reservoir excitons with totally polarized electron spins, $n_X \approx 1.8 \times 10^9$~cm$^{-2}$. Such a density can be easily created in our experiments. Indeed, the number of absorbed photons per excitation pulse, $n_{phot}$, is calculated using the well-known expressions for the absorption coefficient~\cite{Ivchenko-book}, $\eta = (2\Gamma_0\Gamma_{NR})/[(\tilde{\omega}_0 - \omega)^2 + (\Gamma_0+\Gamma_{NR})^2]$. Taking into account the spectral overlap of the pump laser pulses and the Xhh resonance, we obtain for the experimental conditions of Fig.~\ref{Fig2}: $n_{phot} \approx 2 \times 10^{11}$ photons per pulse per cm$^2$. Only a small fraction of bright excitons created by the absorbed photons can be scattered from the light cone into the nonraditative reservoir, $f \sim \tau_X/\tau_{ac} \sim 0.1$ where $\tau_X \approx 10$~ps is the characteristic time of the exciton radiative recombination and $\tau_{ac} \sim 100$~ps is the exciton-phonon scattering time~\cite{Trifonov-PRB2015}. Finally the areal density of excitons in the nonradiative reservoir, $n_{X} \sim 2 \times 10^{10}$~cm$^{-2}$,  is still 10 times larger than the required exciton density. Taking into account the possible loss of the electron spin polarization during the exciton scaterring, we obtain the energy shift comparable with the experimentally observed one. 

\begin{figure}[t]
\centering
\includegraphics[width=1\linewidth]{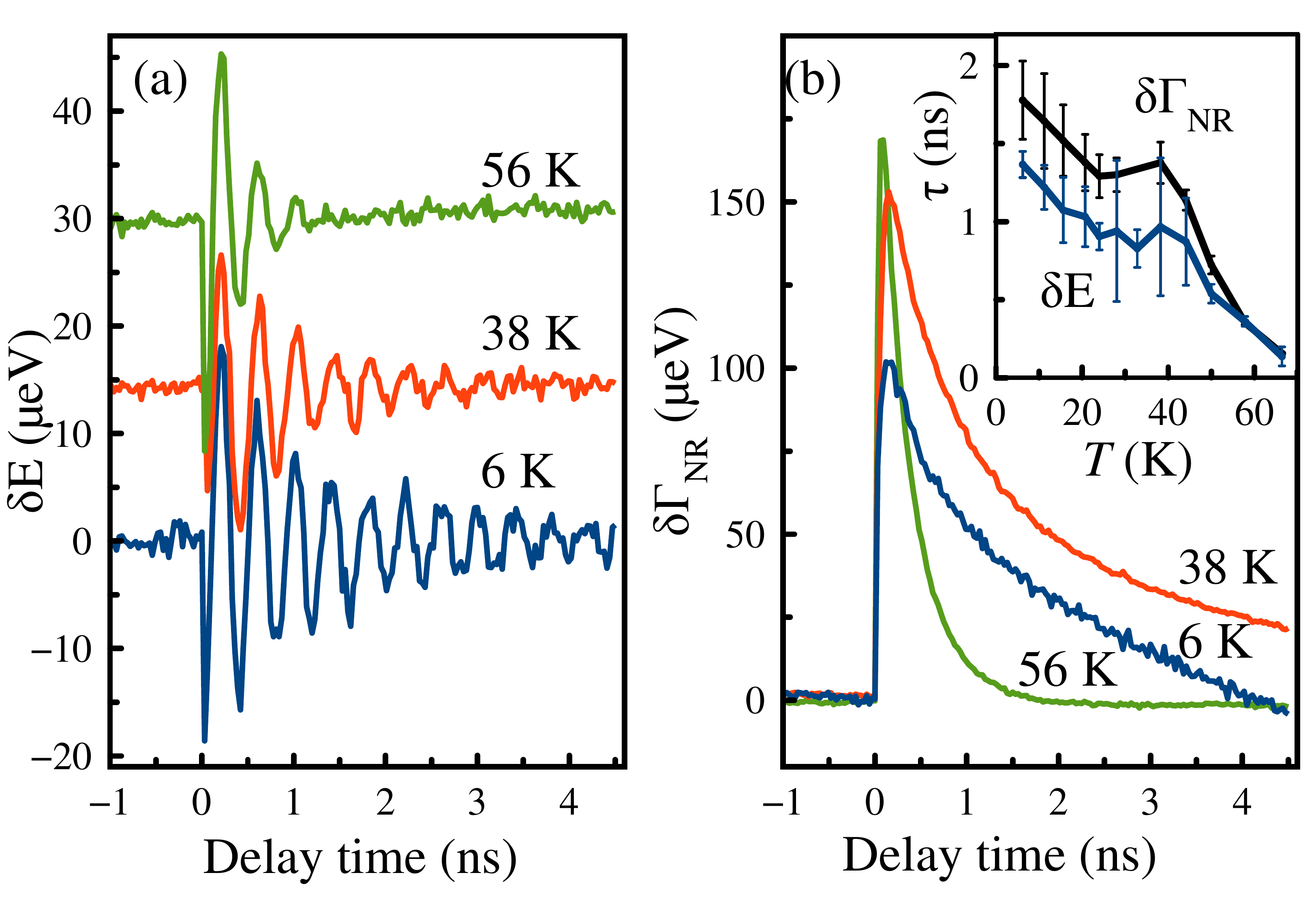} 
\caption{The pump-probe time delay dependence of the exciton energy splitting $\delta E_{hh}$  (a) and of the nonradiative broadening $\hbar \delta \Gamma_{NR} (t)$ (b) at different temperatures. The curves in (a) are shifted for clarity; in (b) only the single pump pulse induced broadening is shown. Inset in (b) shows the temperature dependence of the characteristic decay times of the oscillations and of the nonradiative broadening. The excitation density is $P = 15$~W/cm$^2$.}
\label{Fig3}
\end{figure}

To further support this conclusion, we have measured the dynamics of the exciton energy splitting $\delta E(t)$ at different temperatures and compared it with the dynamics of the nonradiative broadening of the Xhh resonance, $\hbar \delta \Gamma_{NR} (t) = \hbar \Gamma_{NR} (t) - \hbar \Gamma_{NR} (t_m)$, where $t_m$ is a small negative delay time. A representative set of these data is shown in Fig.~\ref{Fig3}. A close similarity is observed in the dynamics of both the energy oscillations and the broadening. A phenomenological fit of the dynamics of the exciton energy splitting  by  $\delta E(t) = A_{\delta E} \exp(-t/\tau_{\delta E})\cos{(\omega t +\varphi)}$ and of the broadening  by  $\hbar\delta\Gamma_{NR}(t) = A_{NR}\exp(-t/\tau_{NR})$ allows one to obtain the characteristic decay times $\tau_{\delta E}$ and $\tau_{NR}$ of these processes.  Their dependence on the sample temperature is shown in the inset to Fig.~\ref{Fig3}(b). 

The obtained data clearly demonstrate that $\delta E(t)$ decays in time nearly with the same rate as the nonradiative broadening $\hbar\delta\Gamma_{NR}(t)$. Having in mind that the decay of the broadening is governed by the depopulation of the exciton reservoir~\cite{Trifonov-PRB2015}, we conclude that the decay of the oscillations is also mainly related to the exciton depopulation. The observed small difference in the decay times could be, in principle, explained by electron spin relaxation or dephasing. The effective time of these processes in high-quality QWs is very large~\cite{ColtonPRB2004, Belykh-PRB2016}. 

The analysis above supports our conclusion that the observed behaviour of the oscillating signal is caused by polarized electrons belonging to excitons in the nonradiative reservoir. The spins of these electrons are  coupled with those of holes via the exchange interaction. The magnitude of this interaction, $\Delta_0 \sim 10 - 20$~$\mu$eV, is larger than the Zeeman splittings observed in our experiments, $\delta E_Z = g \mu_B B \approx 11$~$\mu$eV at $B = 0.5$~T. In the presence of the exchange interaction, the magnetic-field dependence of the oscillation frequency should be nonlinear~\cite{Oestreich-PRB1996}, which would contradict to the experimental observation, see the inset in Fig.~\ref{Fig2}.

We assume that the exchange interaction is effectively switched off in the nonradiative excitons~\cite{Dyakonov-book}. During their relatively long lifetime the hole spin can be lost due to the spin-orbit interaction because the thermal energy $kT \gg \Delta_0$. In fact, due to the interaction with the phonon bath, the hole spin orientation can be changed many times during one period of the electron spin precession in the external magnetic field. A characteristic time of the hole spin relaxation is of order of several tens of ps~\cite{Oestreich-PRB1996, Dyakonov-book}. Even when the hole spin polarisation is lost, there are fluctuations of electron-hole exchange interaction which might destroy the electron spin polarisation~\cite{Dyakonov-PRB1997}. However, in our case the exchange interaction is much weaker ($\Delta_0< 20 \mu$eV) then exciton-collision-induced broadening ($\Gamma_{NR}>100\mu$eV). During the period of electron spin precession in the fluctuating field the exciton is scattered many times by other excitons, so that the fluctuating field is effectively averaged. This process is similar to the well known motional narrowing~\cite{MotionalNarowing}.

In conclusion, we have directly observed the exciton energy shift caused by the exchange interaction of the photocreated excitons with those in the nonradiative reservoir in high-quality QWs. The shift oscillates in time when the transverse magnetic field is applied to the structure. The oscillations decay on a nano-second time scale, which is orders of magnitude longer than exciton radiative lifetime. We attribute the oscillations to the spin precession of electrons belonging to the nonradiative (dark) excitons. Our experiment clearly shows that the electron-hole exchange interaction in dark excitons is suppressed due to depolarization of holes during the large lifetime of these excitons. We have theoretically modeled the exchange interaction of electrons in the photocreated excitons with those in the nonradiative excitons and obtain relevant interaction constants. 

The authors are grateful to I. A. Yugova for fruitful discussions. The authors acknowledge SPbU
for a research grant 11.34.2.2012 and the Russian-German collaboration in the frame of ICRC TRR 160 project supported by the RFBR grant 10-52-12019. A.V.T. acknowledges the RFBR grant 18-32-00516. I.~V.~I. acknowledges the RFBR grant 16-02-00245 a. The SPbU resource center "Nanophotonics" is acknowledged for the structure studied in the present paper.


%
%
%
\pagebreak
\onecolumngrid
\appendix
\center{\Huge{Appendix}}

{\huge{Exchange interaction of the bright excitons with reservoir excitons and free electrons in quantum wells}}

\section{Exchage-interaction-induced energy splitting of exciton states}

The spin-hamiltonian of the heavy-hole exciton interacting with electron and holes reads:
\begin{equation}
\hat{H}_S = \Delta_0 \hat{i}_z \hat{s}_z + J_{hh} n_h \hat{i}_z \langle I_z \rangle + J_{ee} n_e \left ( \hat{\vec{s}} \cdot \langle \vec{S} \rangle\right)+ J_{eh} \left( n_e \hat{i}_z \langle S_z \rangle + n_h \hat{s}_z  \langle I_z \rangle \right).
\label{Ham1}
\end{equation}
Here $\hat{\vec{s}}$ and $\hat{\vec{i}}$ are the spin operators of the electron and the hole comprising the exciton, $\langle \vec{S} \rangle$ and $\langle \vec{I} \rangle$ are the mean values of spins of the electrons and holes in the reservoir, and $n_e$ and $n_h$ are their areal densities. The constants $\Delta_0$, $J_{hh}$, $J_{ee}$, and $J_{eh}$ characterize, respectively, the exchange interaction of the electron and hole in the exciton, the exciton hole with a reservoir hole, the exciton electron with a reservoir electron, and the exciton electron (hole) with a reservoir hole (electron). The heavy hole spin in GaAs-based quantum wells (QWs) is characterized by a very anisotropic $g$-factor tensor with nearly zero in-plane component. Therefore only the $z$-projection of the hole spin is included in the Hamiltonian~(\ref{Ham1}). There are no terms in this Hamiltonian mixing the hole spin states $i_z = +3/2$ and $i_z = -3/2$, therefore it can be separated in two independent Hamiltonians: 
\begin{eqnarray}
\hat{H}_{S+} = \frac{3}{2} \Delta_0 \hat{s}_z + \frac{3}{2} J_{hh} n_h \langle I_z \rangle + J_{ee} n_e \left ( \hat{\vec{s}} \cdot \langle \vec{S} \rangle   \right ) + J_{eh} \left( \frac{3}{2} n_e \langle S_z \rangle + \hat{s}_z \langle I_z \rangle  \right), \nonumber \\
\hat{H}_{S-} = -\frac{3}{2} \Delta_0 \hat{s}_z - \frac{3}{2} J_{hh} n_h \langle I_z \rangle + J_{ee} n_e \left ( \hat{\vec{s}} \cdot \langle \vec{S} \rangle   \right ) + J_{eh} \left( \frac{3}{2} n_e \langle S_z \rangle + \hat{s}_z \langle I_z \rangle  \right).
\label{Ham2}
\end{eqnarray}
In what follows we take into account that the electron-hole exchange interaction is much smaller than the electron-electron and hole-hole ones, $J_{eh} \ll J_{ee}$,~$J_{hh}$~\cite{Dyakonov-book}. Besides, we assume that the hole concentration in the reservoir is negligible small, $n_h \approx 0$. Therefore we save only the first and third terms in the Hamiltonians $\hat{H}_{S+}$ and $\hat{H}_{S-}$.  

Solution of the Schr\"{o}dinger equation with these Hamiltonians gives rise to the following energies of the heavy-hole exciton states: 
\begin{eqnarray}
E_{b+} = \frac{1}{2}  \sqrt{\left(\frac{3}{2} \Delta_0 + J_{ee} n_e  \langle S_z \rangle\right)^2 +  \left( J_{ee} n_e  \langle S_y \rangle \right)^2}, \nonumber   \\
E_{d+} = - \frac{1}{2}  \sqrt{\left(\frac{3}{2} \Delta_0 + J_{ee} n_e  \langle S_z \rangle\right)^2 +  \left( J_{ee} n_e  \langle S_y \rangle \right)^2}, \nonumber   \\
E_{b-} = \frac{1}{2}  \sqrt{\left(\frac{3}{2} \Delta_0 - J_{ee} n_e  \langle S_z \rangle\right)^2 +  \left( J_{ee} n_e  \langle S_y \rangle \right)^2}, \nonumber   \\
E_{d-} = -\frac{1}{2}  \sqrt{\left(\frac{3}{2} \Delta_0 - J_{ee} n_e  \langle S_z \rangle\right)^2 +  \left( J_{ee} n_e  \langle S_y \rangle \right)^2}. 
 \label{energies}
\end{eqnarray}
Here we assume that the optical excitation creates a spin polarization of reservoir electrons along the $z$ axis. An external magnetic field applied along the $x$ axis, $B = B_x$, can turn the polarization towards the $y$ axis. Besides, we assume the external magnetic field to be small enough so that no valuable spin polarization is created along the $x$ axis.

At zero spin polarization of the reservoir electrons, $\langle S_z \rangle = \langle S_y \rangle = 0$, the levels $E_{b+}$ and $E_{b-}$ are degenerate and correspond to the bright excitons, the degenerate levels $E_{d+}$ and $E_{d-}$ correspond to the dark excitons. When the reservoir electrons are polarized along the $z$ axis, $\langle S \rangle = \langle S_z \rangle$, the bright and dark exciton states are not mixed and their energies linearly depend on $J_{ee} n_e  \langle S_z \rangle$. This interaction affects the excitons as an effective magnetic field giving rise to the Zeeman splitting of the bright exciton levels $E_{b+}$ and $E_{b-}$ as well as the dark exciton levels $E_{d+}$ and $E_{d-}$. When the spin polarization of the reservoir electrons is turned perpendicular to the $z$ axis by the external magnetic field, $\langle S \rangle = \langle S_y \rangle$, the bright and dark exciton states are mixed and their energies nonlinearly depend on $J_{ee} n_e  \langle S_z \rangle$.

The intensities of optical transitions to the exciton states can be described by the general expression: 
\begin{equation}
I = \frac{2}{3}I_0 \left ( \frac{3}{4} - j_z s_z \right ).
\label{intensity}
\end{equation}
In the absence of the reservoir spin polarization, $j_z = \pm 3/2$, $s_z = \mp 1/2$ for the bright exciton and the intensities of optical transitions in the $\sigma^+$ and $\sigma^-$ polarizations, $I_+ = I_- = I_0$. For the dark exciton, $j_z = \pm 3/2$, $s_z = \pm 1/2$ and the intensities are zero. 

The intensity $I_0$ is proportional to the constant $\Gamma_0$ of the exciton radiative damping. The integral of the resonant exciton reflection, in the general case, nonlinearly depends on $\Gamma_0$, see Eqs.~(1, 2) in the main text of the paper. In our case, however, the nonradiative damping constant, $\Gamma_{NR}$, is considerably larger than $\Gamma_0$. Therefore expression~(2) of the main text can be expanded in a series over $\Gamma_0$:
\begin{equation}
R(\omega) = |r_s|^2 + \delta R(\omega) \Gamma_0 + \cdots,
\end{equation}
in which the first frequency-dependent term is linear in $\Gamma_0$.

In the presence of the reservoir spin polarization, the quantization axis for the electron spin of the radiative exciton is directed along the total effective magnetic field of the exchange interaction, $\vec{B}_{\text{eff}} = (3/2)\Delta_0 \vec{i}_z + J_{ee} n_e \langle \vec{S} \rangle$, where $\vec{i}_z$ is the unit vector along the $z$ axis. Here we neglect the external magnetic field, which is small relative to $\vec{B}_{\text{eff}}$ in the experiments under discussion. Correspondingly, the projection of electron spin, $s_z$, is determined by the expression:
\begin{equation}
s_z = \frac{1}{2}\frac{\frac{3}{2}\Delta_0 \pm J_{ee} n_e \langle S_z \rangle}{\sqrt{\left(\frac{3}{2} \Delta_0 \pm J_{ee} n_e  \langle S_z \rangle\right)^2 +  \left( J_{ee} n_e  \langle S_y \rangle \right)^2}}.
\label{sz}
\end{equation}
Here the sign ``+'' is for the positive projection of vector $\langle \vec{S} \rangle$ on the $z$ axis and sign ``-'' is for the negative one.

From Eqs.~(\ref{intensity}, \ref{sz}) we obtain for the intensities of optical transitions to the all four levels:
\begin{eqnarray}
I_{b+} = \frac{I_0}{2} \left ( 1 + \frac{\frac{3}{2}\Delta_0 + J_{ee} n_e \langle S_z \rangle}{ \sqrt{\left(\frac{3}{2} \Delta_0 + J_{ee} n_e  \langle S_z \rangle\right)^2 +  \left( J_{ee} n_e  \langle S_y \rangle \right)^2}} \right ), \nonumber \\
I_{d+} = \frac{I_0}{2} \left ( 1 - \frac{\frac{3}{2}\Delta_0 + J_{ee} n_e \langle S_z \rangle}{ \sqrt{\left(\frac{3}{2} \Delta_0 + J_{ee} n_e  \langle S_z \rangle\right)^2 +  \left( J_{ee} n_e  \langle S_y \rangle \right)^2}} \right ), \nonumber \\
I_{b-} = \frac{I_0}{2} \left ( 1 + \frac{\frac{3}{2}\Delta_0 - J_{ee} n_e \langle S_z \rangle}{ \sqrt{\left(\frac{3}{2} \Delta_0 - J_{ee} n_e  \langle S_z \rangle\right)^2 +  \left( J_{ee} n_e  \langle S_y \rangle \right)^2}} \right ), \nonumber \\
I_{d-} = \frac{I_0}{2} \left ( 1 - \frac{\frac{3}{2}\Delta_0 - J_{ee} n_e \langle S_z \rangle}{ \sqrt{\left(\frac{3}{2} \Delta_0 - J_{ee} n_e  \langle S_z \rangle\right)^2 +  \left( J_{ee} n_e  \langle S_y \rangle \right)^2}} \right ).  
\end{eqnarray}
It follows from these expressions that, when the effective magnetic field is directed along the $z$ axis, $\langle S \rangle = \langle S_z \rangle$, only the intensities of the bright exciton transitions are not zero. When the effective field has an arbitrary orientation, all four transitions are allowed due to the bright-dark exciton states mixing. The mixing magnitude depends on the ratio of  $(3/2)\Delta_0$ and $J_{ee} n_e \langle S \rangle$. 

In the heterostructure under study, the exchange constant $\Delta_0<20$~$\mu$eV (see the last section of these materials). The magnitude of the exchange interaction with reservoir electrons can be considerably larger because the splitting of bright exciton states, $E_{b+} - E_{b-}$, reaches 100~$\mu$eV, see Fig.~2 of the main text. This means that the bright and dark exciton states can be strongly mixed when $\langle \vec{S} \rangle$ is perpendicular to the $z$ axis. Correspondingly, optical transitions to the initially dark exciton states should be, in principle, observable in this case. However, the large nonradiative broadening of the exciton resonances (hundreds of $\mu$eV) does not allow one to separate these resonances. We, therefore, calculate the average energies of exciton transitions in the $\sigma^+$ and $\sigma^-$ circular polarizations: 
\begin{eqnarray}
\langle E_+ \rangle = \frac{E_{b+}I_{b+} + E_{d+} I_{d+}}{I_{b+}+I_{d+}} = +\frac{3}{4}\Delta_0 + \frac{1}{2}J_{ee} n_e \langle S_z \rangle, \nonumber \\
\langle E_- \rangle = \frac{E_{b-}I_{b-} + E_{d-} I_{d-}}{I_{b-}+I_{d-}} = +\frac{3}{4}\Delta_0 - \frac{1}{2}J_{ee} n_e \langle S_z \rangle. 
\end{eqnarray}
It follows from these expressions that the exciton energy splitting, $\delta E = \langle E_+ \rangle - \langle E_- \rangle$, observed experimentally is described by the simple expression: 
\begin{equation}
\delta E = J_{ee} n_e  \langle S_z\rangle,
\label{deltaE}
\end{equation}
that is directly determined by the exchange interaction with the reservoir electrons.

A similar analysis can be done for the case of exchange interaction of the radiative excitons with the polarized reservoir excitons. If we assume a fast hole spin relaxation in the reservoir excitons, the interaction is determined by the exchange of spins of the electrons comprising the radiative and reservoir excitons. In this case all above expressions become valid for the analysis of the exchange interaction and the difference of these two cases is described by the exchange constant $J_{ee}$.

\section{Exchange constants}

The general expression for the exchange constant  for interaction of two electrons reads~\cite{LandayVol3}:
\begin{equation}
J = S_n \iint  \frac{e^2}{\varepsilon |\vec{r}_1 -\vec{r}_2|} \Psi_1(\vec{r}_1)\Psi^{*}_1(\vec{r}_2)\Psi_2(\vec{r}_2) \Psi^{*}_2(\vec{r}_1) \,d^3r_1\,d^3r_2.
\label{obmen_int}
\end{equation}
Here $\Psi_1(\vec{r})$, $\Psi_2(\vec{r})$ are the wave functions of the electrons and $\vec{r_1}$, $\vec{r_2}$ are their coordinates; $\varepsilon$ is the dielectric constant of the medium. The quantity $S_n$ is the normalizing area. Its physical meaning is evident from the relation: $S_n = 1/n_e$, that is, $S_n$ is the average area per one electron. 

For the case of exchange interaction of the electron in an exciton and of a free electron, the general expression~(\ref{obmen_int}) is transformed to:
\begin{eqnarray}
J^{xe}_{ee} &=& S_n \iiint \iiint  \iiint \frac{e^2}{\varepsilon \sqrt{(z_{e}-z_{f})^2 + |\vec{\rho}_{e} - \vec{\rho}_{f}|^2}} \nonumber  \\ 
&\times& \Psi_f(z_{e}, \vec{\rho}_{e}) \Psi^{*}_{ex} (z_{e}, z_h, \vec{\rho}_{e}, \vec{\rho}_h) \Psi^{*}_f(z_{f}, \vec{\rho}_{f}) \Psi_{ex} (z_{f}, z_h, \vec{\rho}_{f}, \vec{\rho}_h) \nonumber \\
&\times&  \,dz_{e} \,dz_{f} \,dz_{h} \,d^2\rho_{e} \,d^2\rho_{f} \,d^2\rho_{h} ,
\label{Jxe}
\end{eqnarray}
where $\Psi_f(z_f, \vec{\rho}_f)$ is the wave function of the free electron and $\Psi_{ex}(z_e, z_h,\vec{\rho}_{e}, \vec{\rho}_h)$ is the exciton wave function. Both functions are written in the cylindrical coordinate system to take into account the symmetry of the problem. The variables $z_f, \vec{\rho}_f$ and $z_e, \vec{\rho}_e$ describe the coordinates of electrons in the reservoir and in the exciton, respectively; $z_h, \vec{\rho}_h$ are the coordinates of the hole in the exciton.

The interaction of two excitons via the exchange by electron spins is characterized by the exchange integral~\cite{Ciuti-PRB1998}: 
\begin{eqnarray}
J^{xx}_{ee} &=& S_n \iiint \iiint \iiint  \iiint \frac{e^2}{\varepsilon \sqrt{(z_{e1}-z_{e2})^2 + |\vec{\rho}_{e1} - \vec{\rho}_{e2}|^2}} \nonumber  \\ 
&\times& \Psi_{ex}(z_{e1}, z_{h1}, \vec{\rho}_{e1},\vec{\rho}_{h1}) \Psi_{ex}(z_{e2}, z_{h2}, \vec{\rho}_{e2},\vec{\rho}_{h2}) \nonumber \\
&\times& \Psi_{ex}^{*}(z_{e2}, z_{h1}, \vec{\rho}_{e2},\vec{\rho}_{h1}) \Psi_{ex}^{*}(z_{e1}, z_{h2}, \vec{\rho}_{e1},\vec{\rho}_{h2}) \nonumber \\
&\times&  \,dz_{e1} \,dz_{h1}  \,dz_{e2} \,dz_{h2} \,d^2\rho_{e1} \,d^2\rho_{h1} \,d^2\rho_{e2} \,d^2\rho_{h2}.
 \label{Jxx}
\end{eqnarray}
It is assumed here that there is no motion of the excitons along the QW plane, that is, the wave vector $K_{ex} =0$. It is really small for the excitons created by optical pumping with nearly normal incidence of light. It is also small for the reservoir excitons, $K_{ex} \ll 1/a_B$, for the liquid helium temperatures used in the experiments. Here $a_B \approx 14$~nm is the exciton Bohr radius. 

\section{Calculation of the exchange constants}

\subsection{Numerical calculations of the exchange constants}

Direct numerical calculations of $J_{ee}^{xe}$ and $J_{ee}^{xx}$ are performed in two steps. On the first step, the exciton wave function, $\Psi_{ex}(z_{e}, z_{h}, \vec{\rho}_{e},\vec{\rho}_{h})$, is obtained by direct numerical solution of the three-dimensional Schr\"{o}dinger equation for an exciton in the GaAs QW under study~\cite{Khramtsov-JAP2016}. The wave function of a free electron along the $z$ axis in the QW is calculated by numerical solution of the one-dimensional Schr\"{o}dinger equation. Along the $x$ and $y$ axes the function is assumed to be a constant determined from normalization over the area $S_n$. Both the exciton and electron wave functions are assumed to be real functions, that is, $\Psi = \Psi^*$. 

The numerically obtained exciton wave function, $\Psi_{ex}(z_{e}, z_{h}, \vec{\rho})$, depends on the relative electron-hole distance in the QW plane, $\vec{\rho} = \vec{\rho}_{e}-\vec{\rho}_{h}$, due to the assumed cylindrical symmetry of the problem. Therefore the total exciton wave function is transformed to:
\begin{equation}
 \Psi_{ex}(z_{e}, z_{h}, \vec{\rho}_e, \vec{\rho}_h) = \frac{1}{\sqrt{S_n}}\Psi_{ex}(z_{e}, z_{h}, \vec{\rho}).
 \end{equation}
 The cylindrical symmetry allows one to simplify expressions~(\ref{Jxe}, \ref{Jxx}):
\begin{eqnarray}
J^{xe}_{ee} &=& S_n \int \iiint  \iiint \frac{e^2}{\varepsilon \sqrt{(z_{e}-z_{f})^2 + |\vec{\rho}_{e} - \vec{\rho}_{f}|^2}} \nonumber  \\ 
&\times& \Psi_f(z_{e}, \vec{\rho}_{e}) \Psi_{ex} (z_{e}, z_h, \vec{\rho}_{e}) \Psi_f(z_{f}, \vec{\rho}_{f}) \Psi_{ex} (z_{f}, z_h, \vec{\rho}_{f}) \nonumber \\
&\times&  \,dz_{e} \,dz_{f} \,dz_{h} \,d^2\rho_{e} \,d^2\rho_{f} ,
\label{Jxe_s}
\end{eqnarray}
\begin{eqnarray}
J^{xx}_{ee} &=& \int \iiint \iiint  \iiint \frac{e^2}{\varepsilon \sqrt{(z_{e1}-z_{e2})^2 + |\vec{\rho}_{e1} - \vec{\rho}_{e2}|^2}} \nonumber  \\ 
&\times& \Psi_{ex}(z_{e1}, z_{h1}, \vec{\rho}_{e1}) \Psi_{ex}(z_{e2}, z_{h2}, \vec{\rho}_{e2}-\vec{\rho}_{h}) \nonumber \\
&\times& \Psi_{ex}(z_{e2}, z_{h1}, \vec{\rho}_{e2}) \Psi_{ex}(z_{e1}, z_{h2}, \vec{\rho}_{e1}-\vec{\rho}_{h}) \nonumber \\
&\times&  \,dz_{e1} \,dz_{h1}  \,dz_{e2} \,dz_{h2} \,d^2\rho_{e1} \,d^2\rho_{h} \,d^2\rho_{e2},
 \label{Jxx_s}
\end{eqnarray}
It is assumed in Eqs.~(\ref{Jxe_s}, \ref{Jxx_s}) that the numerically obtained exciton wave functions are normalized to unity.

On the second step, the exchange integrals~(\ref{Jxe_s}) and (\ref{Jxx_s}) are calculated using a simple Monte Carlo method. Using a pseudorandom number generator, coordinates of electrons and holes are generated in a large enough three-dimensional region where the electron and exciton wave functions are noticeable nonzero: $z_e, z_h \in [-50, 50]$~nm, $\rho \in (0, 120]$~nm. Then the expressions under integrals $J_{xe}$ and $J_{xx}$ are calculated and accumulated. Typically $10^7 - 10^8$ random coordinates have been used to obtain the integral with reasonable accuracy. The results of the calculations are given in Tab.~\ref{TabI}.

\subsection{Approximate calculations of the exchange constants}

The direct numerical solution of the Schr\"{o}dinger equations, in particular for an exciton in a QW, is a complex problem. Therefore it is useful to compare the numerical results with those obtained in different approximations of the wave functions. First we consider the widely used approximations applicable for narrow QWs~\cite{Ivchenko-book}. The quasi-two-dimensional electron wave function reads:
\begin{equation}
\Psi_f (z_f, \vec{\rho}_f)= \frac{1}{\sqrt{S_n}}\sqrt{\frac{2}{L_z}}\cos(\frac{\pi z_f}{L_z}),
\label{electron_waveFunc}
\end{equation}
where $L_z$ is the QW width. 
The exciton wave function is chosen as:
\begin{equation}
\Psi_{ex} (z_e, z_h, \vec{\rho}_e, \vec{\rho}_h) = \frac{1}{\sqrt{S_n}} \frac{2}{L_z} \cos(\frac{\pi z_e}{L_z}) \cos(\frac{\pi z_h}{L_z})  \psi_r (\vec{\rho}_e-\vec{\rho}_h). 
\label{exciton_waveFunc}
\end{equation}
We assume that the wave function of the relative motion is given by: 
\begin{equation}
\psi_r (\vec{\rho}) =\frac{1}{a_{B}} \sqrt{\frac{2}{\pi}} \exp\left(-\frac{|\vec{\rho}|}{a_{B}}\right),
\label{psi-rho} 
\end{equation}
which is valid for two-dimensional excitons~\cite{Ivchenko-book}. Here $a_{B}$ is the exciton Bohr radius along the QW plane.

In relatively narrow QWs with low-height barriers, the electron and exciton wave functions may noticeably penetrate into the barriers. To take into account this possible effect, we also have calculated the exchange constants using the numerically obtained electron and hole wave functions, $\Psi_{e(h)}^{\text{num}}(z_{e(h)})$, and the exciton wave function of type:
\begin{equation}
 \Psi_{ex} (z_e, z_h, \vec{\rho}_e, \vec{\rho}_h) = \Psi_e^{\text{num}}(z_e) \Psi_h^{\text{num}}(z_h) \psi_r (\vec{\rho}_e-\vec{\rho}_h). 
 \label{seminum}
 \end{equation} 

Substituting functions~(\ref{electron_waveFunc}, \ref{exciton_waveFunc}) into Eq.~(\ref{Jxe}), we obtain:
\begin{eqnarray}
J^{xe}_{ee} &=& \frac{1}{S_n} \left(\frac{2}{L_z}\right)^3 \iiint\iiint\iiint \frac{e^2}{\varepsilon \sqrt{(z_{e1}-z_{e2})^2 + |\vec{\rho}_{e1} - \vec{\rho}_{e2}|^2}}  \nonumber  \\
&\times& \cos^2(\frac{\pi z_{e1}}{L_z}) \cos^2(\frac{\pi z_{e2}}{L_z}) \cos^2(\frac{\pi z_{h}}{L_z}) \psi_r (\vec{\rho}_{e1} - \vec{\rho}_{h}) \psi_r (\vec{\rho}_{e2} - \vec{\rho}_{h}) \,dz_{e1} \,dz_{e2} \,dz_{h} \,d^2\rho_{e1} \,d^2\rho_{e2} \,d^2\rho_{h} \nonumber \\
&=& \left(\frac{2}{L_z}\right)^2 \iiint\iiint \frac{e^2}{\varepsilon \sqrt{(z_{e1}-z_{e2})^2 + |\vec{\rho}_{e1} - \vec{\rho}_{e2}|^2}} \nonumber \\
&\times& \cos^2(\frac{\pi z_{e1}}{L_z}) \cos^2(\frac{\pi z_{e2}}{L_z}) \psi_r (\vec{\rho}_{e1}) \psi_r (\vec{\rho}_{e2}) \,dz_{e1} \,dz_{e2} \,d^2\rho_{e1} \,d^2\rho_{e2}.
\label{electron_int2}
\end{eqnarray}

Similarly the expression for the constant of the exciton-exciton exchange interaction~(\ref{Jxx_s}) with model function~(\ref{exciton_waveFunc}) can be obtained:
\begin{eqnarray}
J^{xx}_{ee} &=& \left(\frac{2}{L_z}\right)^2 \iint\iiint\iiint \frac{e^2}{\varepsilon \sqrt{(z_{e1}-z_{e2})^2 + |\vec{\rho}_{e1} - \vec{\rho}_{e2}|^2}} \cos^2(\frac{\pi z_{e1}}{L_z}) \cos^2(\frac{\pi z_{e2}}{L_z}) \nonumber  \\
&\times& \psi_r (\vec{\rho}_{e1})\psi_r (\vec{\rho}_{e2})\psi_r (\vec{\rho}_{e1} - \vec{\rho}_{h}) \psi_r (\vec{\rho}_{e2} - \vec{\rho}_{h}) \,dz_{e1} \,dz_{e2} \,d^2\rho_{e1} \,d^2\rho_{e2} \,d^2\rho_{h}.
\label{Jxx-model}
\end{eqnarray}

The exchange constants $J_{ee}^{xe}$ and  $J_{ee}^{xx}$ are calculated by integration of expressions~(\ref{Jxe_s}, \ref{Jxx_s}) and (\ref{electron_int2}, \ref{Jxx-model}) by the Monte Carlo method. Two cases are considered. In the first case, the penetration of  the electron and exciton wave functions into the barrier layers due to their relatively small height is taken into account by the use of the numerical electron function $\Psi_e^{\text{num}}(z_e)$ and of the exciton wave function~(\ref{seminum}) in the general expressions~(\ref{Jxe_s}, \ref{Jxx_s}). 
In the second case, the barriers are assumed to be infinitely high, no penetration is possible, and functions~(\ref{electron_waveFunc}) and (\ref{exciton_waveFunc}) are used in the calculations of intergals~(\ref{electron_int2}, \ref{Jxx-model}). The value of the exciton Bohr radius, $a_{B} = 12.6$~nm, is obtained from the fit of the numerical exciton wave function in the middle of the QW at $z_e = z_h$ by function~(\ref{psi-rho}) along the QW plane. The results for both the cases are given in Tab.~\ref{TabI}.

\begin{table}[h]
\caption{Exchange constants obtained in various approximations: ``Numerical'' is the calculation with use of the numerically obtained electron and exciton wave functions; ``Model~(\ref{seminum})'' is  use of the numerically obtained electron wave function and of the exciton function~(\ref{seminum}); ``Infinite barriers'' is use of functions~(\ref{electron_waveFunc}, \ref{exciton_waveFunc}).} 
\begin{center}
\begin{tabular}{|c|c|c|}
\hline
Exciton wave function & $J^{xx}_{ee}$ ($\mu eV\mu m^2$) &  $J^{xe}_{ee}$($\mu eV\mu m^2$)\\
\hline
Numerical &  11.4 $\pm$ 0.9 & 18 $\pm$ 2\\
\hline
Model~(\ref{seminum}) & 9.3 $\pm$ 0.7  & 17.9 $\pm$ 0.6\\
\hline
Infinite barriers  & 10.3 $\pm$ 0.1  & 19.5 $\pm$ 0.2 \\
\hline
\end{tabular}
\end{center}
\label{TabI}
\end{table}

As seen from the table, all the calculations give rise to very similar results: $J_{ee}^{xx} \approx 10$~$\mu{\text{eV}}\cdot \mu{\text{m}}^2$; $J_{ee}^{xe} \approx 18$~$\mu{\text{eV}}\cdot \mu{\text{m}}^2$, independent on the approximation used. In particular, these results are almost insensitive to the penetration of the electron and exciton wave functions into the barriers. The overlap of the free electron and exciton wave functions plays the main role in the exchange interaction and it is only slightly changed when penetration takes place. The main result of these calculation is that, for the relatively narrow QW, the model functions~(\ref{electron_waveFunc}), (\ref{exciton_waveFunc}) and the approximation of infinitely high barriers can be used to obtain reliable values of the exchange constants with appropriate accuracy. 

\subsection{Analytical estimate of $J_{ee}^{xe}$}

Expression~(\ref{electron_int2}) can be rewritten in terms of the Coulomb energy of some fictitious charge distributions. Indeed:
\begin{eqnarray}
J^{xe}_{ee} &=& 16 \pi a_{B}^2 \cdot \frac{1}{2}\iiint\iiint \frac{v(z_e, \vec{\rho}_e) v(z_f, \vec{\rho}_f)}{\varepsilon \sqrt{(z_{e}-z_{f})^2 + |\vec{\rho}_{e} - \vec{\rho}_{f}|^2}} \,dz_{e} \,dz_{f} \,d^2\rho_{e} \,d^2\rho_{f} \nonumber \\
 &=& 16 \pi a_{B}^2 \cdot V_C.
\label{Jxe-model}
\end{eqnarray}
Here $V_C$ is the Coulomb energy of a charge system described by the charge density:
\begin{equation}
v(z, \vec{\rho}) = \frac{e \cos^2(\frac{\pi z}{L_z}) e^{-|\vec{\rho}|/a_{B}}}{L_z \pi a_{B}^2}.
\end{equation}

To estimate the Coulomb energy, we approximate the charge distribution by an uniformly charged oblate ellipsoid of revolution with small axis, $a_1 = L_z/2$, and large axis, $a_2 = 2 a_{B}$. It is shown in the textbook~\cite{LandayVol2} that this problem can be reduced to the calculation of the energy of a uniformly charged ball. The result is:
\begin{equation}
V_C (a_{B}, \nu) = \frac{3e^2}{5\varepsilon a_{B}} f(\nu),
\label{VC}
\end{equation}
where $\nu = a_1/a_2$, $f(\nu) = 1/\sqrt{1-\nu^2} \arctan(\sqrt{(1-\nu^2)/\nu^2})$. 
The value $\nu=1$ corresponds to the spherically symmetric charge distribution, at which $f(\nu) =1$. At $\nu = 0 $ (the strongly oblate ball), $f(\nu) = \pi/2$. As seen the variation in the values of $f(\nu)$ is not large.
From Eqs.~(\ref{Jxe-model}, \ref{VC}) we obtain:
\begin{equation}
J^{xe}_{ee} = \frac{48 \pi e^2}{5 \varepsilon a_{B}} f(\nu) a^2_{B}. 
\label{Vex}
\end{equation}

We have verified the result~(\ref{Vex}) by the direct numerical calculation of the constant~$J^{xe}_{ee}$ with the  ellipsoidal charge distribution:
\begin{eqnarray}
J^{xe}_{ee} =  8 \pi a_{B}^2 \iiint \iiint \frac{e^2}{\varepsilon\sqrt{(z_e-z_f)^2+(\vec{\rho}_e-\vec{\rho}_f)^2}} \times \nonumber \\
\phi(z_e,\vec{\rho}_e) \phi(z_f,\vec{\rho}_f)  \,dz_f \,d^2\vec{\rho}_f \,dz_e \,d^2\vec{\rho}_e,
\label{Vex1}
\end{eqnarray}
where
\begin{equation}
\phi_{ex}(z_e,\vec{\rho}_e-\vec{\rho}_h) =  \frac{3}{\pi a_{B}^2 L_z}
    \left\{
    \begin{matrix}
    1, & \quad \mbox{if } \frac{z_e^2}{(L_z/4)^2} + \frac{(\rho_e-\rho_h)^2}{a_B^2} \le 1 \\
    0, & \quad \mbox{if } \frac{z_e^2}{(L_z/4)^2} + \frac{(\rho_e-\rho_h)^2}{a_B^2} > 1.
    \end{matrix} \right.
    \label{ellipsoid}
\end{equation}
The calculations have shown that the numerically obtained result with use of Eqs.~(\ref{Vex1}, \ref{ellipsoid}) precisely coincides with that obtained from Eq.~(\ref{Vex}): $J^{xe}_{ee} =  29.3$~$\mu$eV~$\mu$m$^{2}$. The calculation parameters are: $a_{B} = 12.6 $~nm, $\varepsilon = 12.53$.
 
The obtained value of the exchange constant $J^{xe}_{ee}$ slightly exceeds those presented in Tab.~\ref{TabI}. Taking into account the roughness of approximation of the the charge distribution by expression~(\ref{ellipsoid}), the discrepancy between these values can be considered as appropriate. 

\section{Evaluation of the constant $\Delta_0$}

The exchange splitting between the bright and dark exciton states, $(3/2) \Delta_0$, is too small in many cases to be directly measured in the experiment. In bulk GaAs, $(3/2) \Delta_0^{\text{bulk}} < 10$~$\mu$eV~\cite{Blackwood1994}. It increases in QWs due to the stronger overlap of the electron and hole wave functions. This effect can be taken into account by introducing the enhancement factor~\cite{Blackwood1994}:
\begin{equation}
F = \frac{\int | \Phi_{\text{QW}}(z)|^2 dz}{\int | \Phi_{\text{bulk}}(z)|^2 dz}.
\end{equation}
Here $\Phi(z)$ is the cross-section of the exciton wave function $\Psi(z_e, z_h, \rho)$ taken at $z_e = z_h = z$ and $\rho = 0$, that is, at the coinciding electron and hole coordinates.

We have calculated this enhancement factor by  use of the numerically obtained exciton wave functions for the QW under study and for a wide enough, bulk-like, QW ($L_z = 200$~nm), in which the exchange interaction almost coincides with that for bulk GaAs. The obtained value is: $F = 1.8$. This means that the exchange splitting in the structure under study, $(3/2) \Delta_0^{\text{QW}} < 20$~$\mu$eV. It is small compared to the energy splitting caused by the interaction with polarized electron spins in the reservoir.



\begin{thebibliography}{99}
\bibitem{comment2} We should note that there are also dark excitons with small in-plane wave vectors but with spin $S = \pm 2$ which block the interaction with photons possessing spin $S = \pm 1$. The lifetime of the dark excitons, however, is much smaller than that of the excitons in the nonradiative reservoir because the energy splitting between  bright and dark excitons is small compared to the thermal energy $kT$ even at helium temperatures. Therefore the dark excitons in quantum wells can be quite rapidly converted into  bright ones and emit light.
\end{thebibliography}
\end{document}